\documentclass[8.5pt,twoside,twocolumn]{article}
\usepackage{amsmath}
\usepackage[super,sort&compress,comma]{natbib}
\usepackage{mhchem}
\usepackage{times,mathptmx}
\usepackage{sectsty}
\usepackage{balance}

\usepackage{graphicx} %eps figures can be used instead
\usepackage{lastpage}
\usepackage[format=plain,justification=raggedright,singlelinecheck=false,font=small,labelfont=bf,labelsep=space]{caption}
\usepackage{fancyhdr}

\usepackage{subfig}
\usepackage{SIunits}
\usepackage[compact]{titlesec}
\begin{document}

\thispagestyle{plain}
\fancypagestyle{plain}{
%\fancyhead[L]{\includegraphics[height=8pt]{headers/LH}}
%\fancyhead[C]{\hspace{-1cm}\includegraphics[height=20pt]{headers/CH}}
%\fancyhead[R]{\includegraphics[height=10pt]{headers/RH}\vspace{-0.2cm}}
\renewcommand{\headrulewidth}{1pt}}
\renewcommand{\thefootnote}{\fnsymbol{footnote}}
\renewcommand\footnoterule{\vspace*{1pt}%
\hrule width 3.4in height 0.4pt \vspace*{5pt}}
\setcounter{secnumdepth}{5}

\makeatletter
\def\subsubsection{\@startsection{subsubsection}{3}{10pt}{-1.25ex plus -1ex minus -.1ex}{0ex plus 0ex}{\normalsize\bf}}
\def\paragraph{\@startsection{paragraph}{4}{10pt}{-1.25ex plus -1ex minus -.1ex}{0ex plus 0ex}{\normalsize\textit}}
\renewcommand\@biblabel[1]{#1}
\renewcommand\@makefntext[1]%
{\noindent\makebox[0pt][r]{\@thefnmark\,}#1}
\makeatother
\renewcommand{\figurename}{\small{Fig.}~}
\sectionfont{\large}
\subsectionfont{\normalsize}

\fancyfoot{}
%\fancyfoot[LO,RE]{\vspace{-7pt}\includegraphics[height=9pt]{headers/LF}}
%\fancyfoot[CO]{\vspace{-7.2pt}\hspace{12.2cm}\includegraphics{headers/RF}}
%\fancyfoot[CE]{\vspace{-7.5pt}\hspace{-13.5cm}\includegraphics{headers/RF}}
\fancyfoot[RO]{\footnotesize{\sffamily{1--\pageref{LastPage} ~\textbar  \hspace{2pt}\thepage}}}
\fancyfoot[LE]{\footnotesize{\sffamily{\thepage~\textbar\hspace{3.45cm} 1--\pageref{LastPage}}}}
\fancyhead{}
\renewcommand{\headrulewidth}{1pt}
\renewcommand{\footrulewidth}{1pt}
\setlength{\arrayrulewidth}{1pt}
\setlength{\columnsep}{6.5mm}
\setlength\bibsep{1pt}

\twocolumn[
  \begin{@twocolumnfalse}
\noindent\LARGE{\textbf{Active modulation of electromagnetically induced transparency analogue in terahertz hybrid metal-graphene metamaterials}}
\vspace{0.6cm}

\noindent\large{\textbf{Shuyuan Xiao,\textit{$^{a}$} Tao Wang,$^{\ast}$ \textit{$^{a}$} Tingting Liu,\textit{$^{b}$} Xicheng Yan,\textit{$^{a}$} Zhong Li,\textit{$^{c,d}$} and Chen Xu\textit{$^{e}$}}}\vspace{0.5cm}

%Please note that \ast indicates the corresponding author(s) but no footnote text is required.

\noindent\textit{\small{\textbf{Received Xth XXXXXXXXXX 20XX, Accepted Xth XXXXXXXXX 20XX\newline
First published on the web Xth XXXXXXXXXX 200X}}}

\noindent \textbf{\small{DOI: 10.1039/b000000x}}
\vspace{0.6cm}
%Please do not change this text.

\noindent \normalsize{Metamaterial analogues of electromagnetically induced transparency (EIT) have been intensively studied and widely employed for slow light and enhanced nonlinear effects. In particular, the active modulation of the EIT analogue and well-controlled group delay in metamaterials have shown great prospects in optical communication networks. Previous studies have focused on the optical control of the EIT analogue by integrating the photoactive materials into the unit cell, however, the response time is limited by the recovery time of the excited carriers in these bulk materials. Graphene has recently emerged as an exceptional optoelectronic material. It shows an ultrafast relaxation time on the order of picosecond and its conductivity can be tuned via manipulating the Fermi energy. Here we integrate a monolayer graphene into metal-based terahertz (THz) metamaterials, and realize a complete modulation in the resonance strength of the EIT analogue at the accessible Fermi energy. The physical mechanism lies in the active tuning the damping rate of the dark mode resonator through the recombination effect of the conductive graphene. Note that the monolayer morphology in our work is easier to fabricate and manipulate than isolated fashion. This work presents a novel modulation strategy of the EIT analogue in the hybrid metamaterials, and pave the way towards designing very compact slow light devices to meet future demand of ultrafast optical signal processing.}
\vspace{0.5cm}
 \end{@twocolumnfalse}
  ]

\section{Introduction}\label{sec1}
%Footnotes
%\footnotetext{\dag~Electronic Supplementary Information (ESI) available: [details of any supplementary information available should be included here]. See DOI: 10.1039/b000000x/}

%Please use \dag to cite the ESI in the main text of the article.
%If you article does not have ESI please remove the the \dag symbol from the title and the above footnotetext.

\footnotetext{\textit{$^{a}$~Wuhan National Laboratory for Optoelectronics, Huazhong University of Science and Technology, Wuhan 430074, People's Republic of China. E-mail: wangtao@hust.edu.cn}}
\footnotetext{\textit{$^{b}$~School of Electronic Information and Communications, Huazhong University of Science and Technology, Wuhan 430074, People's Republic of China}}
\footnotetext{\textit{$^{c}$~Center for Materials for Information Technology, The University of Alabama, Tuscaloosa 35487, United State of America}}
\footnotetext{\textit{$^{d}$~Department of Physics and Astronomy, The University of Alabama, Tuscaloosa 35487, United State of America}}
\footnotetext{\textit{$^{e}$~Department of Physics, New Mexico State University, Las Cruces 88001, United State of America}}
%\footnotetext{\textit{$^{b}$~Address, Address, Town, Country. }}

%additional addresses can be cited as above using the lower-case letters, c, d, e... If all authors are from the same address, no letter is required

%\footnotetext{\ddag~Additional footnotes to the title and authors can be included \emph{e.g.}\ `Present address:' or `These authors contributed equally to this work' as above using the symbols: \ddag, \textsection, and \P. Please place the appropriate symbol next to the author's name and include a \texttt{\textbackslash footnotetext} entry in the the correct place in the list.}
Electromagnetically-induced transparency (EIT) refers to the formation of a narrow transparency window within a broad absorption profile upon the application of an indirect excitation and the quantum destructive interference between the two excitation channels in a three-level atomic system.\cite{boller1991observation,harris1997electromagnetically,fleischhauer2005electromagnetically} This phenomenon is always accompanied by an extreme modification of the dispersion properties and thus potentially useful in many applications, such as slow light\cite{hau1999light,lukin2001controlling,longdell2005stopped,novikova2012electromagnetically} and enhanced nonlinear effects.\cite{schmidt1996giant,wu2003large,zhang2007opening} However, realizing the conventional quantum EIT requires the stable optical pumping and often cryogenic temperature, whose complexities severely restrict practical applications, especially with respect to on-chip integration. These barriers have been overcome by considering that the underlying physics behind the EIT phenomenon is actually classical, and the analogue behaviors can be reproduced using coupled harmonic oscillators and RLC electric circuits.\cite{garrido2002classical,souza2015electromagnetically} This physical insight leads to the realization of the EIT analogues in a series of classical optical systems, such as coupled microresonators,\cite{smith2004coupled,xu2006experimental,peng2007optical} photonic crystal waveguides,\cite{mingaleev2008coupled,yang2009all} a waveguide side-coupled to resonators\cite{yanik2004stopping,lu2012plasmonic} and metamaterials\cite{zhang2008plasmon,papasimakis2008metamaterial,liu2009plasmonic}, which are robust and free from the scathing experimental requirements of quantum optics. In particular, the metamaterial analogues of EIT through the near field coupling between the bright and dark mode resonators, have enabled the realization of this phenomenon at frequencies in radio-frequency (RF),\cite{tassin2012electromagnetically,han2014engineering,li2015low,zhu2015magnetic} terahertz (THz),\cite{singh2009coupling,kim2010multi,singh2011sharp,singh2011observing,liu2012electromagnetically,manjappa2017magnetic} near-infrared\cite{liu2013multispectral,liu2014robust,si2014plasmon,yang2014all} and visible regimes\cite{biswas2013plasmon,yan2015directional,mun2016polarization} through defining a correspondingly tailored geometry for the unit cell. Due to the subwavelength thickness, these EIT analogues with the accompanying slow light and enhanced nonlinear effects have shown great prospects in designing very compact devices, such as optical filters,\cite{hu2015tailoring,cheng2015ultrathin} optical buffers\cite{nakanishi2013storage,wang2013novel} and ultrasensitive biosensors.\cite{liu2009planar,verellen2011plasmon,liu2017high}

For practical applications, an active modulation of the EIT analogue and consequently the well-controlled group delay are highly desirable. To this end, an active metamaterial is a superior candidate. Up to now, there are two main strategies to actively modulate the EIT analogues in metamaterial devices. One is to integrate the metamaterials with the photoactive materials, such as Silicon (Si), and the active modulation of EIT can be realized through the change in the damping rate of the dark mode resonator by the increasing conductivity of Si under optical pumping.\cite{chen2008experimental,gu2012active,xu2016frequency,manjappa2017active} However, the response time in this kind of modulation process is limited by the recovery time of the excited carriers in Si ($\sim 1$ ms), which severely hinders applications in ultrafast optoelectronics. On the other hand, graphene, a two-dimensional (2D) material with a plethora of exceptional electronic and photonic properties, has garnered enormous attention.\cite{novoselov2004electric,bonaccorso2010graphene,de2013graphene} The relaxation time of the excited carriers in the monolayer graphene is on the order of picosecond, showing a promising future for ultrafast response.\cite{rana2009carrier,yee2011ultrafast,li2014ultrafast} Moreover, the conductivity of graphene can also be continuously tuned via manipulating its Fermi energy by electric gating or photo-induced doping, which lays the direct foundation for efficient real-time control of resonance in metamaterials.\cite{ju2011graphene,he2015graphene,he2015tunable,fan2015tunable,fan2016electrically,linder2016graphene} Therefore, a variety of graphene metamaterials have been proposed to realize the EIT analogues in the recent decade.\cite{shi2013plasmonic,cheng2013dynamically,ding2014tuneable,luo2016flexible,tang2016dynamic,zhao2016graphene,xia2016dynamically,he2016terahertz} Nevertheless, the isolated graphene resonator in the unit cell can not be expediently tuned in practice and the ultrasmall feature size that corresponds to resonance entering into the THz gap is also challenging in the nanoscale fabrication.

In the present work, we propose, for the first time to the best of our knowledge, an active modulation of the EIT analogue in THz resonant metamaterials through integrating a monolayer graphene into the unit cell. The simulation results show that a complete modulation in the resonance strength of the EIT analogue can be realized at the accessible Fermi energy of graphene. The theoretical analysis based on the coupled harmonic oscillator model and distributions of the electric field and surface charge density reveal that the active modulation is attributable to the change in the damping rate of the dark mode resonator by the recombination effect of the conductive graphene. In addition, the well-controlled group delay accompanying EIT is also calculated for slow light applications. The picosecond-order response time of graphene facilitates the ultrafast optical modulation, and the monolayer morphology possesses the advantage of being easier to fabricate and manipulate, showing much better efficiency and feasibility than previous studies. Therefore, this work not only demonstrate the use of graphene in the THz hybrid metamaterials to the active modulation of the EIT analogues, but also pave the way towards designing very compact slow light devices with ultrafast response, which may play a vital role in the future THz communications.

\section{The geometric structure and numerical model}\label{sec2}
The schematic illustration and geometric parameters of our proposed hybrid structure are shown in Fig. 1(a) and (b). The unit cell of Aluminum (Al)-based resonant metamaterials is arranged in a periodical array with lattice constants of $P_{x}=80$ $\micro\meter$ and $P_{y}=120$ $\micro\meter$, and composed of a cut wire (CW) and a pair of identical but oppositely oriented split ring resonators (SRRs) on the top of a Si substrate. The CW is $L=85$ $\micro\meter$ in length and $W=5$ $\micro\meter$ in width; the SRRs are $l=29$ $\micro\meter$ in side length and $g=5$ $\micro\meter$ in split gap. The coupling distance between the CW and the SRRs is set to $s=7$ $\micro\meter$. The thickness of both of the resonators are $t_{Al}=200$ nm and the substrate is assumed to be semi-infinite. The optical properties of Al in the THz regime are described by the Drude model\cite{ordal1985optical}
\begin{equation}
    \varepsilon_{Al}=\varepsilon_{\infty}-\frac{\omega_{p}^{2}}{\omega^{2}+i\omega\gamma},\label{eq1}
\end{equation}
where the plasma frequency $\omega_{p}=2.24\times 10^{16}$ rad/s and the damping constant $\gamma=1.22\times 10^{14}$ rad/s. The refractive index of Si is taken as $n_{Si}=3.42$.
\begin{figure*}[ht]
\centering
\includegraphics[scale=0.8]{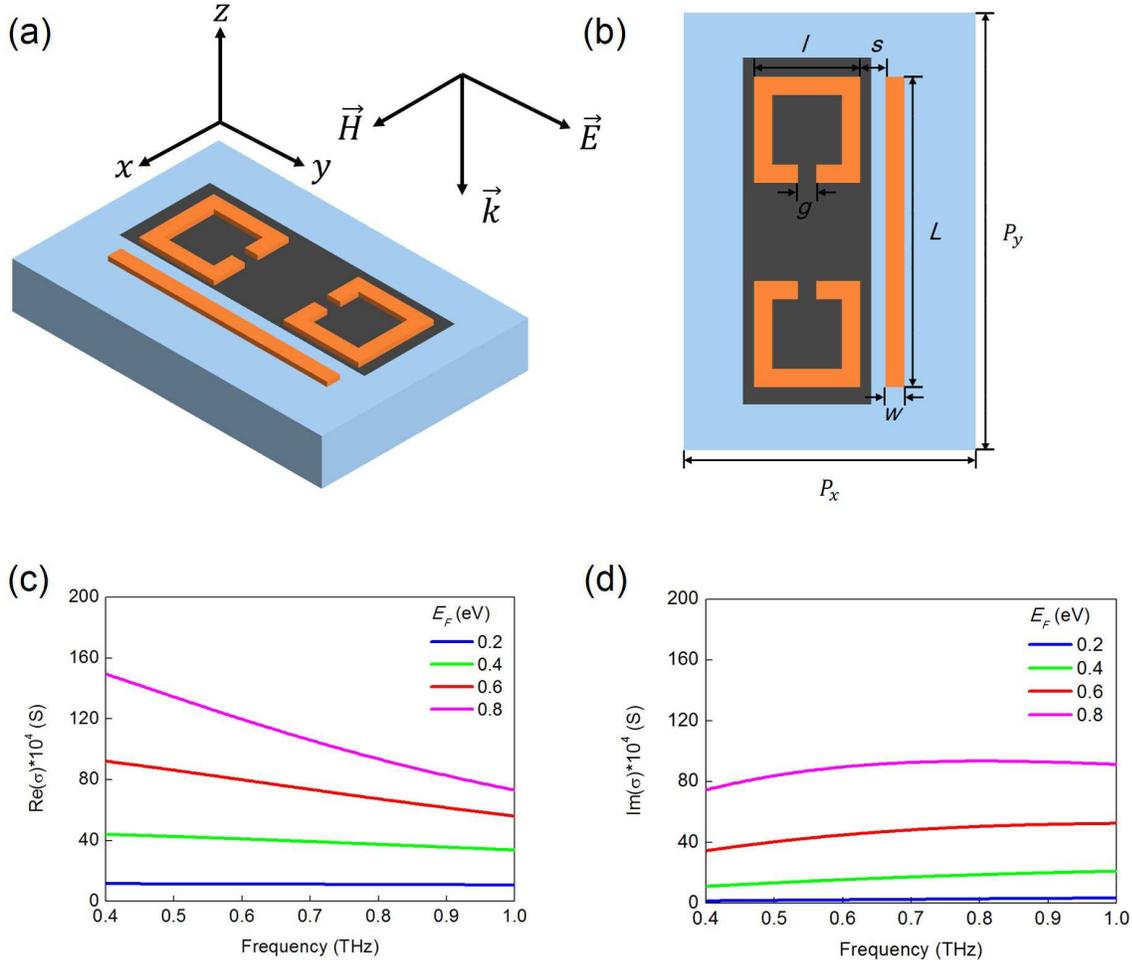}
\caption{(a) The schematic illustration of our proposed hybrid metamaterials with a normally incident plane wave. (b) The top view of the unit cell. The geometrical parameters are $L=85$ $\micro\meter$, $W=5$ $\micro\meter$, $l=29$ $\micro\meter$, $g=5$ $\micro\meter$, $s=7$ $\micro\meter$, $P_{x}=80$ $\micro\meter$ and $P_{y}=120$ $\micro\meter$, respectively. (c) The real part and (d) the imaginary part of the frequency dependent graphene conductivity. The Fermi energy is varied from 0.2 eV to 0.8 eV, as shown in the insets.\label{fig:1}}
\end{figure*}

The monolayer graphene is placed under the SRRs and modeled as a 2D flat sheet. The graphene conductivity can be derived using the random-phase approximation (RPA) in the local limit, including both the intraband and interband processes\cite{zhang2014coherent,zhang2015towards,zhang2016towards}
\begin{equation}
    \begin{split}
      \sigma_{g} &=\sigma_{intra}+\sigma_{inter}=\frac{2e^{2}k_{B}T}{\pi\hbar^{2}}\frac{i}{\omega+i\tau^{-1}}\ln[2\cosh(\frac{E_{F}}{2k_{B}T})]\\
      &+\frac{e^2}{4\hbar}[\frac{1}{2}+\frac{1}{\pi}\arctan(\frac{\hbar\omega-2E_{F}}{2k_{B}T}) \\
      &-\frac{i}{2\pi}\ln\frac{(\hbar\omega+2E_{F})^{2}}{(\hbar\omega-2E_{F})^{2}+4(k_{B}T)^{2}}],\label{eq2}
    \end{split}
\end{equation}
where $e$ is the charge of an electron, $k_B$ is the Boltzmann constant, $T$ is the operation temperature, $\hbar$ is the reduced Planck's constant, $\omega$ is the angular frequency of the incident light, $\tau$ is the carrier relaxation time and $E_F$ is the Fermi energy, respectively. In the lower THz regime, the optical losses originated from the interband process is negligible due to the Pauli exclusion principle and therefore graphene conductivity can be safely reduced to the Drude-like model\cite{li2016monolayer,xiao2016tunable,xiao2017strong}
\begin{equation}
    \sigma_{g}=\frac{e^{2}E_{F}}{\pi\hbar^{2}}\frac{i}{\omega+i\tau^{-1}},\label{eq3}
\end{equation}
where the carrier relaxation time $\tau=(\mu E_F)/(e v_F^2)$ depends on the carrier mobility $\mu$, the Fermi energy $E_{F}$ and the Fermi velocity $v_{F}$. Here we employ $\mu=3000$ cm$^{2}$/V$\cdot$s and $v_{F}=1.1\times 10^{6}$ m/s throughout the calculations, which are consistent with the experimental measurements.\cite{zhang2005experimental,jnawali2013observation} Consequently, the frequency dependent graphene conductivity at the selected values of the Fermi energy are shown in Fig. 1(c) and (d).

The full-wave numerical simulations are performed using the finite-difference time-domain method (FDTD Solutions, Lumerical Inc., Canada). In the calculations, the moderate mesh grid is adopted to balance the simulation time and accuracy. The periodical boundary conditions are employed in the $x$ and $y$ directions and perfectly matched layers are utilized in the $z$ direction along the propagation of the incident plane wave.

\section{Simulation results and discussions}\label{sec3}
In order to investigate the EIT analogue in the Al-based resonant metamaterials, we simulate three sets of arrays with the unit cell composed of a CW, a pair of SRRs and both of them, respectively, and the corresponding transmission spectra are shown in Fig. 2. The CW exhibits a typical localized surface plasmon (LSP) resonance mode at 0.65 THz, which is the basic and direct response of metamaterials to the incident plane wave and possesses a broad resonance line width caused by the strong radiative losses. However, the SRRs that can support an LC resonance mode with a narrow line width at the very similar frequency (not shown) does not respond to this excitation due to the reserved structural symmetry with respect to the polarization of the incident plane wave. When both of the CW and the SRRs are simultaneously arranged into the unit cell, a narrow transparency window is clearly observed within a broad absorption profile at 0.65 THz. In this configuration, the CW that can be directly excited by the incident plane wave acts as a bright mode resonator, and the SRRs that can not be directly excited by this excitation act as a dark mode resonator. The near field coupling between them leads to an indirect excitation of the LC resonance mode in the SRRs and the classical destructive interference between the LSP and LC resonance modes give rise to an EIT analogue.
\begin{figure}[htbp]
\centering
\includegraphics[scale=0.4]{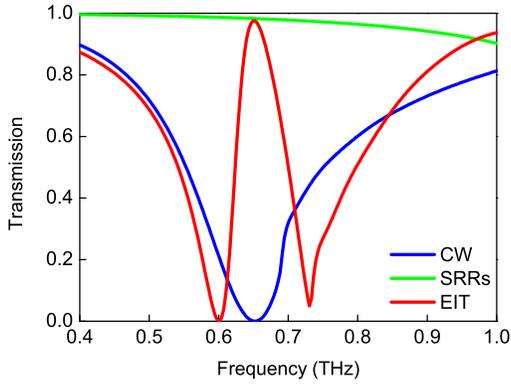}
\caption{The simulated transmission spectra of the CW, the SRRs and the EIT metamaterials.\label{fig:2}}
\end{figure}

We also plot the electric field and surface charge density distributions at the resonance frequency to further illuminate the underlying physics behind the EIT analogue. As shown in Fig. 3(a), before coupling with the SRRs, the CW is directly excited by the incident plane wave with a very strong enhancement of the electric field concentrating on the edges and corners, and correspondingly in Fig. 4(a), a symmetric distribution of the opposite charges along the CW can be nicely observed and the net induced dipoles show a clear tendency to follow the polarization of the incidence plane wave. These distributions correspond to the LSP resonance mode in the CW and validate its function as the bright mode resonator. After the SRRs arranged into the unit cell, however, the strong electric field enhancement entirely transfers to the vicinity of the splits through the near field coupling with the CW in Fig. 3(b), and a large amount of the opposite charges accumulate at two ends of each split, indicating obvious circulation distributions of the surface charge density along SRRs and making them as magnetic dipoles in Fig. 4(b). These are characteristic behaviors of the LC resonance mode, which demonstrate this dark resonance mode in the SRRs is indirectly excited by coupling with the CW. Because the strength of the indirect excitation is comparable to that of the direct excitation but with a $\pi$ phase difference,\cite{zhang2008plasmon,ye2012mapping} the LSP and LC resonance modes interfere destructively with each other, thus giving rise to the narrow transparency window. Note that the electric field in the CW is almost completely suppressed, which validates the destructive interference at the transparency peak.
\begin{figure}[ht]
\centering
\includegraphics[scale=0.27]{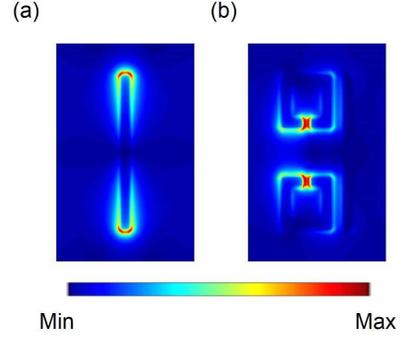}
\caption{The simulated electric field distributions of (a) the CW and (b) the EIT metamaterials at 0.65 THz.\label{fig:3}}
\end{figure}
\begin{figure}[htbp]
\centering
\includegraphics[scale=0.27]{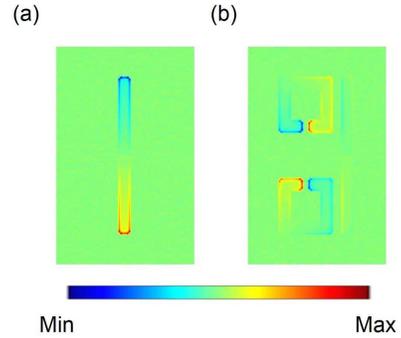}
\caption{The simulated surface charge density distributions of (a) the CW and (b) the EIT metamaterials at 0.65 THz.\label{fig:4}}
\end{figure}

Next, we integrate a monolayer graphene into the unit cell and investigate its modulation effects on the EIT analogue in the Al-based resonant metamaterials. As shown in Fig. 5(a), in the presence of the monolayer graphene under the SRRs, the EIT analogue undergoes a great change in the resonance strength, leading to an on-to-off modulation with the increasing of the Fermi energy of graphene. To quantitatively characterize this induced change in the resonance strength, we introduce the reduction degree in the transmission as $\Delta T=(T_{0}-T_{g})\times100\%$, where $T_{0}$ and $T_{g}$ refer to transmissions at the transparency peak of the EIT analogue without and with the monolayer graphene, respectively. When the Fermi energy starts at 0.2 eV, the transparency peak has an obvious decline from the initial value, and the reduction degree in the transmission is $\Delta T=33\%$. As $E_{F}$ gradually increases to 0.4 eV, the transparency peak continues to fall and $\Delta T$ goes up to $63\%$. Then when $E_{F}$ comes to 0.6 eV, the transparency peak begins to dispear and $\Delta T$ grows to as much as $78\%$. Finally, with the maximum Fermi energy of 0.8 eV, the narrow transparency window merges into the broad absorption profile and only a LSP-like resonance remains with $\Delta T=84\%$, thus obtaining a complete modulation in the resonance strength of the EIT analogue. Here, we would like to highlight that graphene used in the active modulation is in the monolayer morphology rather than isolated fashion, which is easier to fabricate and manipulate. Besides, the on-to-off modulation of the EIT analogue requires only a Fermi energy of 0.8 eV, which has already been accessed by electric gating experimentally.\cite{fang2013gated,fang2013active,hu2015broadly} Therefore, the active modulation of the EIT analogue in our proposed hybrid metamaterials shows much better efficiency and feasibility than previous studies.
\begin{figure}[htbp]
\centering
\includegraphics[scale=0.6]{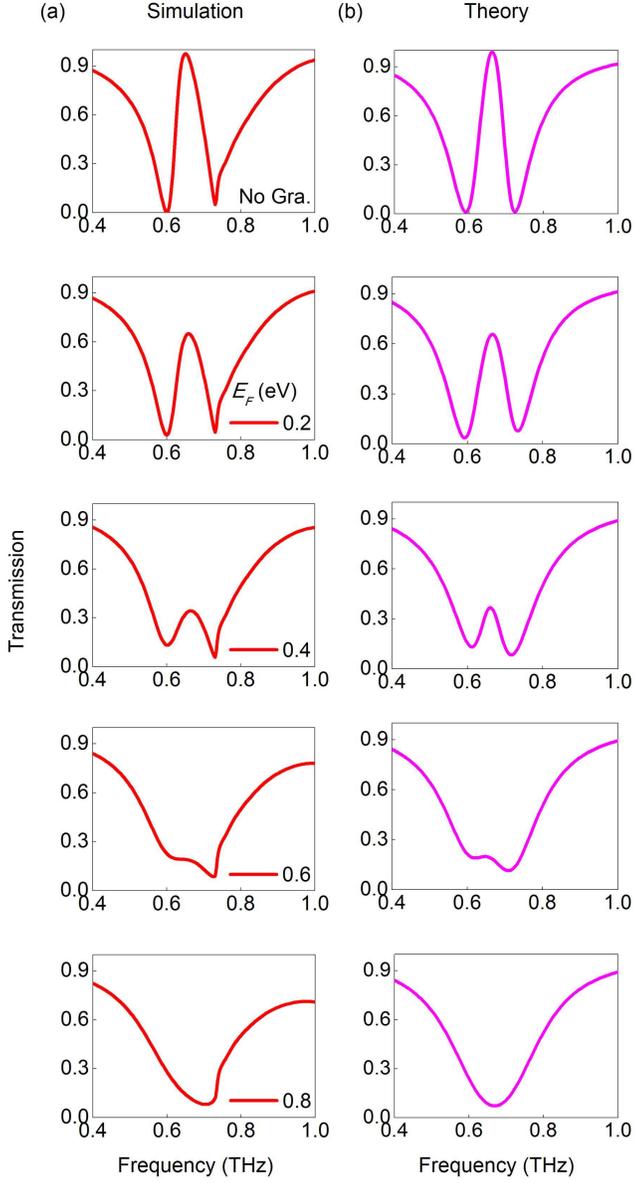}
\caption{\label{fig:5} (a) The simulated transmission spectra of the hybrid metamaterials with the increasing of the Fermi energy of graphene. (b) The corresponding analytical fitted transmission spectra with the increasing of the damping rate of the dark mode resonator.\label{fig:5}}
\end{figure}

In order to reveal the physical mechanism of the active modulation of the EIT analogue, we employ the coupled harmonic oscillator model to quantitatively describe the near field coupling between the bright and dark mode resonators in our proposed hybrid metamaterials. To mimic the quantum EIT in a three-level atomic system, this simple model includes a ground state $|0\rangle$ and two excited states $|1\rangle$ and $|2\rangle$. $|0\rangle\rightarrow|1\rangle$ defines the dipole-allowed transition, corresponding to the LSP bright resonance mode in the CW; $|0\rangle\rightarrow|2\rangle$ defines the dipole-forbidden transition, corresponding to the LC dark resonance mode in the SRRs. Besides, the transition between $|1\rangle$ and $|2\rangle$ corresponds to the near field coupling between the CW and the SRRs, which therefore introduces the destructive interference between two possible excitation channels: $|0\rangle\rightarrow|1\rangle$ and $|0\rangle\rightarrow|1\rangle\rightarrow|2\rangle\rightarrow|1\rangle$, leading to the cancellation of absorption at the EIT frequency. The interference in the metamaterials can be analytically described by the coupled differential equations\cite{zhang2008plasmon,liu2009plasmonic}
\begin{eqnarray}
  &&\ddot{x}_{1}+\gamma_{1}\dot{x}_{1}+\omega_{0}^{2}x_{1}+\kappa x_{2}=E,\label{eq4} \\
  &&\ddot{x}_{2}+\gamma_{2}\dot{x}_{2}+(\omega_{0}+\delta)^{2}x_{2}+\kappa x_{1}=0,\label{eq5}
\end{eqnarray}
where $x_{1}$, $x_{2}$, $\gamma_{1}$ and $\gamma_{2}$ are the resonance amplitudes and the damping rates of the bright and dark mode resonators, respectively, $\omega_{0}$ is the resonance frequency of the bright mode resonator before coupling with the dark mode resonator, $\delta$ is the detuning of the resonance frequency of the dark mode resonator from the bright mode resonator, and $\kappa$ denotes the coupling coefficient between the two resonators. After solving the coupled equations (4) and (5) with the approximation $\omega-\omega_{0}\ll\omega_{0}$, we obtain the susceptibility $\chi$ of the EIT metamaterials\cite{ding2014tuneable,luo2016flexible}
\begin{equation}
    \chi=\chi_{r}+i\chi_{i}\propto\frac{(\omega-\omega_{0}-\delta)+i\frac{\gamma_{2}}{2}}{(\omega-\omega_{0}+i\frac{\gamma_{1}}{2})(\omega-\omega_{0}-\delta+i\frac{\gamma_{2}}{2})-\frac{\kappa^{2}}{4}}.\label{eq6}
\end{equation}
Note that the the energy dissipation is proportional to the imaginary part $\chi_{i}$, therefore the transmission $T$ can be calculated through $T=1-g\chi_{i}$, where $g$ is the geometric parameter indicating the coupling strength of the bright mode resonator with the incident electric field $E$.

As shown in Fig. 5(b), we present the analytical fitted transmission spectra according to the coupled harmonic oscillator model, which exhibit a very good agreement with the simulation results. The corresponding fitting parameters as a function of the Fermi energy of graphene are listed in Fig. 6. In the modulation process, $\gamma_{1}$, $\delta$ and $\kappa$ stay roughly constant with the increasing of $E_{F}$, however, the damping rate of the dark mode resonator $\gamma_{2}$ increases by more than two orders of magnitude. Therefore, the theoretical model indicates that the active modulation is attributable to the change in the damping rate of the dark mode resonator. In our proposed hybrid metamaterials, the monolayer graphene behaves like a quasi-metal with the increasing Fermi energy, and once placed under the SRRs, it connects the two ends of each split, enhance the losses in SRRs and thus weaken the destructive interference between the bright and dark resonance modes. With the maximum Fermi energy of 0.8 eV, the losses become too large to sustain the dark resonance mode in the SRRs, which consequently leads to the disappearance of the transparency peak at the EIT frequency.
\begin{figure}[htbp]
\centering
\includegraphics[scale=0.4]{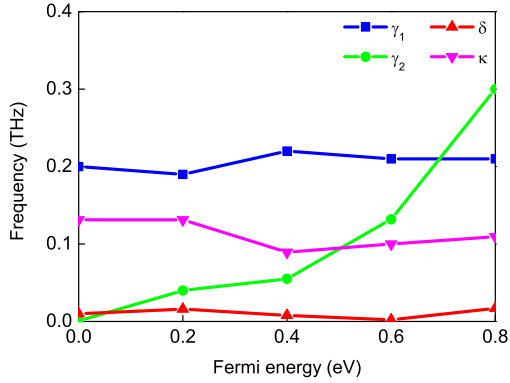}
\caption{The fitting parameters $\gamma_{1}$, $\gamma_{2}$, $\delta$ and $\kappa$ as a function of the Fermi energy of graphene. The unit of $\kappa$ is THz$^{2}$.\label{fig:6}}
\end{figure}

Here we plot the electric field and surface charge density distributions at the resonance frequency to further substantiate the physical mechanism of the active modulation of the EIT analogue. Without the monolayer graphene, the LC resonance mode in the SRRs is indirectly excited by coupling with the CW and in reverse interference destructively with the LSP resonance mode in the CW at the EIT frequency, which reflects in the distributions is the transfer of the strong electric field enhancement from the CW to the SRRs and the accumulation of the opposite charges at two ends of each split, as repeated in Fig. 7(a) and Fig. 8(a). In this configuration, the SRRs exhibit very small damping rate. With the monolayer graphene placed under the SRRs, the opposite charges are recombined and neutralized through this conductive layer, which leads to a strong suppression of the electric field enhancement in the SRRs. This is demonstrated in Fig. 7(b) and Fig. 8(b), when the Fermi energy of graphene gradually increases to 0.4 eV, the electric field is redistributed, where the enhancement in the SRRs significantly declines while that in the CW obviously increases, and correspondingly, the opposite charges become much slighter in the SRRs and the induced dipoles along the CW begin to reappear. With the maximum Fermi energy of 0.8 eV, the recombination effect of the monolayer graphene can be more pronounced with the higher conductivity. As shown in Fig. 7(c) and Fig. 8(c), the opposite charges are almost completely neutralized, and consequently the electric field enhancement in the SRRs is nearly eliminated and that in the CW is fully recovered as before coupling with the SRRs. The SRRs exhibit too large damping rate to support the LC resonance mode, and therefore the destructive interference between the LSP and LC resonance modes disappears.
\begin{figure}[htbp]
\centering
\includegraphics[scale=0.4]{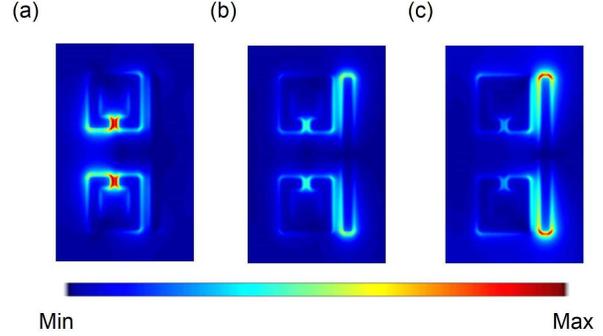}
\caption{The simulated electric field distributions of the EIT metamaterials at 0.65 THz. The corresponding Fermi energies of graphene are (a) no Gra, (b) 0.4 eV, (c) 0.8 eV.\label{fig:7}}
\end{figure}
\begin{figure}[htbp]
\centering
\includegraphics[scale=0.4]{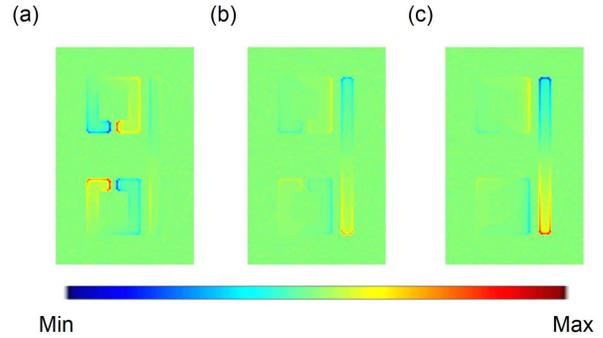}
\caption{The simulated surface charge density distributions of the EIT metamaterials at 0.65 THz. The corresponding Fermi energies of graphene are (a) no Gra, (b) 0.4 eV, (c) 0.8 eV.\label{fig:8}}
\end{figure}

As mentioned above, the EIT phenomenon is always accompanied by the extreme modification of the dispersion properties, which leads to slow light effect. Here we introduce the group delay $t_{g}$ to describe the slow light capability, which is expressed as\cite{lu2012plasmonic}
\begin{equation}
    t_{g}=\frac{d\psi}{d\omega},\label{eq7}
\end{equation}
where $\psi$ is the transmission phase shift. In comparison with the conventionally used group refractive index $n_{g}$, the calculation of $t_{g}$ does not require the effective thickness of the hybrid metamaterials, which is difficult to define due to the semi-infinite substrate, and more importantly, $t_{g}$ is directly related to the practical applications of slow light devices. As shown in Fig. 9(a) and (b), we calculate the transmission phase shift and group delay of the hybrid metamaterials. Without the monolayer graphene, the phase slope is positive and steepest within the transparency window, leading to the largest group delay of 5.72 ps, which corresponds to the group delay of a 1.72-mm distance of free space propagation. With the monolayer graphene placed under the SRRs, the hybrid metamaterials gradually loses its slow light capability with the increasing of the Fermi energy. With the maximum Fermi energy of 0.8 eV, the group delay characteristics of EIT completely turn into those of LSP-like resonance. The well-controlled group delay in our proposed hybrid metamaterials is comparable to previous studies,\cite{gu2012active,xu2016frequency} and can be strategically important in designing very compact slow light devices with ultrafast response.
\begin{figure}[htbp]
\centering
\includegraphics[scale=0.6]{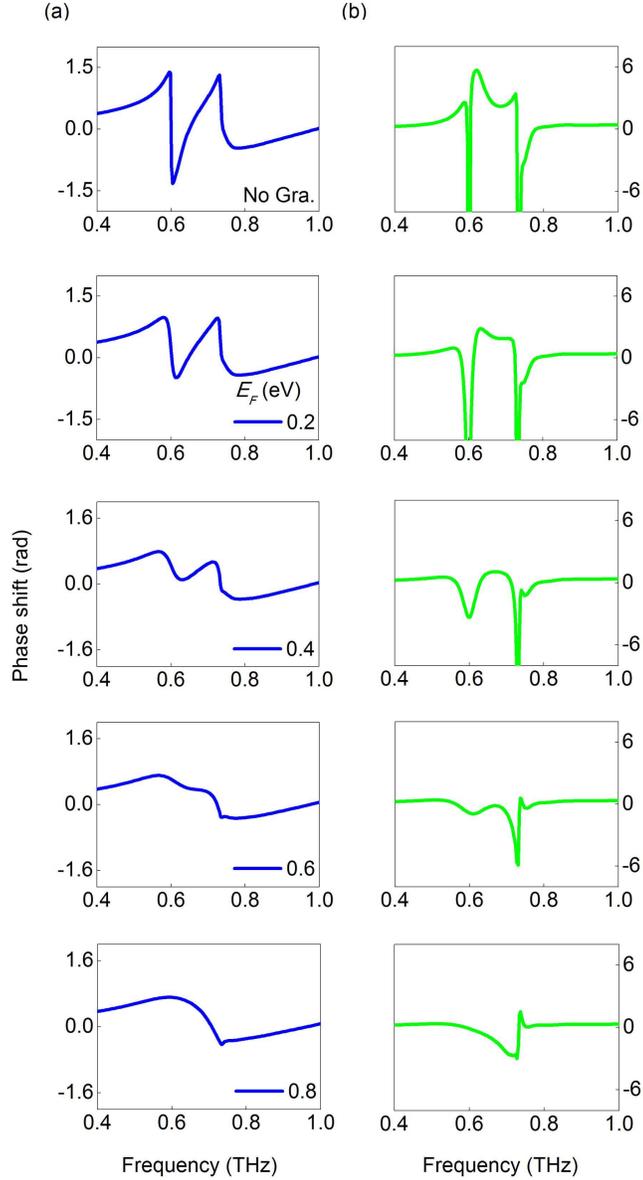}
\caption{\label{fig:5}(a) The simulated transmission phase shift and (b) group delay of the EIT metamaterials with the increasing of the Fermi energy of graphene.\label{fig:9}}
\end{figure}

\section{Conclusions}\label{sec4}
To conclude, we have realized an active modulation of the EIT analogue through integrating a monolayer graphene into the THz resonant metamaterials. The FDTD simulation results show that the on-to-off modulation in the resonance strength of the EIT analogue can be completed with the maximum Fermi energy of 0.8 eV, which has been accessed by electric gating experimentally. We employ the coupled harmonic oscillator model to describe the near field coupling in this hybrid metamaterials that agrees very well with the simulation results, and the theoretical analysis reveals that the active modulation is attributable to the change in the damping rate of the dark mode resonator. The electric field and surface charge density distributions further illuminate that the physical mechanism lies in the recombination effect of the conductive graphene. In addition, the well-controlled group delay is also calculated for slow light applications. The active modulation of the EIT analogue in our proposed hybrid metal-graphene metamaterials opens up great possibilities for designing very compact slow light devices that can find potential applications in THz communication networks.

\section*{Acknowledgments}
The author Shuyuan Xiao (SYXIAO) expresses his deepest gratitude to his Ph.D. advisor Tao Wang for providing guidance during this project. SYXIAO would also like to thank Dr. Qi Lin, Dr. Guidong Liu and Dr. Shengxuan Xia (Hunan University) for beneficial discussions on graphene optical properties. This work is supported by the National Natural Science Foundation of China (Grant No. 61376055, 61006045 and 11647122), the Fundamental Research Funds for the Central Universities (HUST: 2016YXMS024) and the Project of Hubei Provincial Department of Education (Grant No. B2016178).

\balance

%%%REFERENCES%%%
%\bibliography{rsc} %You need to replace "rsc" on this line with the name of your .bib file
\bibliographystyle{rsc} %the RSC's .bst file

\providecommand*{\mcitethebibliography}{\thebibliography}
\csname @ifundefined\endcsname{endmcitethebibliography}
{\let\endmcitethebibliography\endthebibliography}{}

\end{document}